# Dynamic control of mode modulation and spatial multiplexing using hybrid metasurfaces

**Zemeng Lin ,[1] Lingling Huang ,[1, †] Ruizhe Zhao ,[1] Qunshuo Wei ,[1] Thomas Zentgraf,[3] Yongtian Wang ,[1] Xiaowei Li[2, *]**

[1] *School of Optics and Photonics, Beijing Institute of Technology, Beijing, 100081, China*

[2] *Laser Micro/Nano-Fabrication Laboratory, School of Mechanical Engineering, Beijing Institute of Technology, Beijing 100081, China*

[3] *Department of Physics, University of Paderborn, Warburger Straße 100, 33098 Paderborn, Germany*

[†] *huanglingling@bit.edu.cn*

*\*lixiaowei@bit.edu.cn*

**Abstract:** Designing reconfigurable metasurfaces that can dynamically control scattered electromagnetic waves and work in the near-infrared (NIR) and optical regimes remains a challenging task, which is hindered by the static material property and fixed structures. Phase change materials (PCMs) can provide high contrast optical refractive indexes at high frequencies between amorphous and crystal states, therefore are promising as feasible materials for reconfigurable metasurfaces. Here, we propose a hybrid metasurface that can arbitrarily modulate the complex amplitude of incident light with uniform amplitude and full $2\pi$ phase coverage by utilizing composite concentric rings (CCRs) with different ratios of gold and PCMs. Our designed metasurface possesses a bi-functionality that is capable of splitting beams or generating vortex beams by thermal switching between metal and semiconductor states of vanadium oxide ($VO_2$), respectively. It can be easily integrated into low loss photonic circuits with an ultra-small footprint. Our metadevice serves as a novel paradigm for active control of beams, which may open new opportunities for signal processing, memory storage, holography, and anti-counterfeiting.

## 1. Introduction

Metasurfaces have emerged as promising candidates for transforming the interactions between electromagnetic waves and matter [1-4]. By utilizing the arbitrary design freedom of metasurfaces to tailor the amplitude, phase, and polarization response, it might be possible to provide a flexible and compact platform to realize all types of functional devices, such as beam deflector, polarization converter, phase modulator, image processor and so forth [5-12]. Reconfigurability of metasurfaces typically utilize the materials whose optical properties can be modified. By integrating with functional materials such as liquid crystals, graphene (or other 2D materials) and phase change materials (PCMs), metasurfaces can obtain extra freedom of modulating the optical responses. Various reconfigurable mechanisms have been proposed, including mechanical deformation, charge carrier injection, light pumping, thermal modulation, ultra-fast nonlinear all-optical switching, etc. [13-26]. The proposed future applications in their work may bennefit the development of integrated nano-devices and multi-functional metasurfaces. Among those methods, the specific design requirements including working bandwidth, modulation depth, transition condition under external modulations should be carefully taken into account. The free carrier injection, for example, is hard to work in the optical regime for high applied voltages and low modulation depth [27-32]. While other methods may suffer from relatively huge loss and may not be very feasible for the integration of metadevices. Using the PCMs can achieve dynamic functionalities with high efficiency, feasibility, and larger working bandwidth, which makes PCMs a promising choice for reconfigurable metasurfaces.

PCMs, such as germanium antimony telluride (GST), indium antimonide (InSb), gallium lanthanum sulfide (GLS), and vanadium oxide ($VO_2$), are used for data storage, integrated optical circuits, color printing, optical display and so on, by employing their unique reversible, nonvolatile properties [33, 34]. Upon an appropriate stimulus like thermal, optical, or electrical, PCMs offer a flexible control of optical parameters during the phase transition of the composite atom array between crystal and amorphous states or insulator-to-conductor states [35]. Such atomic rearrangements remain stable at room temperature and come with an abrupt change in the physical properties (e.g., complex refractive index and resistivity) after phase transition [36]. This transition can be not only modulated in high speed (some nanoseconds or less) but also reversible with repeated non-volatility (potentially up to $10^{15}$ cycles or more without failure) between amorphous and crystalline states [37]. Though different PCMs could offer tunable optical properties under external excitations, their working wavelength range and phase transition conditions vary from each other. In general, $VO_2$ has the advantage of lower threshold over other PCMs like GST to obtain transition under thermal excitation, which makes it suitable to be integrated into the lower-power consumption devices, such as CMOS chips. The optical properties of $VO_2$ in its amorphous/crystalline state may resemble the ones of the semiconductor and metal property respectively [24, 38]. Also, $VO_2$ can offer the modulation depth as high as 80% [38]. Meanwhile, the reported metasurfaces based on other PCMs mainly work in the mid infrared (MIR) or far infrared (FIR) spectrum which hinders their applications in the near Infrared (NIR). Furthermore, $VO_2$ based metasurfaces are relatively promising to conquer the bandwidth limit. Hence, by integrating $VO_2$ with metasurface, the combination may provide a solution for reconfigurable metadevices.

In previous work, some reconfigurable metasurfaces integrated with PCMs or other methods have been proposed for achieving complex amplitude and polarization modulation of the electromagnetic fields, and also for digital control [39] and other applications such as mimicking synapse [40]. For example, by utilizing periodically arranged GST strips as modulation layer, an active metasurface that can effectively generate tunable ellipticity for arbitrary incident polarization has been achieved [41]. Meanwhile, rewritable focusing devices or reconfigurable holograms based on GST can be achieved by using optical pulses [42]. Also, planar retroflectors using subwavelength-thick reconfigurable C-shaped resonators formed by a liquid metal and an air gap was reported [43]. Though numerous devices employing PCMs emerged, most works are focused on the functions of beam switching, absorber, polarizer in the phase transition process, especially in the NIR regime [44-48]. Furthermore, the previous works may need precise and gradual point-by-point external modulation to accomplish phase transition for achieving the dynamic phenomenon, which are time-consuming [49]. Therefore, by building intelligent, more simple and effective algorithms to achieve active metasurfaces for effective light modulation are highly demanded.

In this paper, we propose and demonstrate a novel type of hybrid reconfigurable metasurface for dynamic control of mode modulation and spatial multiplexing. Such metasurface is composed of composite concentric rings (CCRs) with different ratio of gold and $VO_2$, which can possess dual functionality between the two-phase states. By considering the optical properties of $VO_2$, the CCR operates similarly as a concentric ring in metal (crystalline) state with relatively high conductivity like metal, while functions as a split ring wire loop for $VO_2$ in its semiconductor state. The transition between the two states provides complex amplitude control due to the high contrast refractive index. We delicately design and select eight CCRs to achieve binary phase modulation for the metal (crystalline) state, and fully $2\pi$ phase shift for semiconductor (amorphous) state. The schematic design of such hybrid reconfigurable metasurface is shown in Fig. 1. When illuminated with x polarized light, the output beam with orthogonal polarization shows distinct patterns for different phase state of $VO_2$. At the semiconductor state around room temperature of $VO_2$ (around 25℃), a vortex beam can be generated, as shown in Fig. 1(a). When the metasurface is heated above 85℃ and $VO_2$ transitioned in its metal state [38], as depicted in Fig. 1(b), the designed metasurface functions as the beam splitter and converts the input light into four diffraction lobes (spatial multiplexing). Therefore, by proper design, such metasurface can provide potential applications in fiber communications of the C-band, high-efficiency laser fabrications, signal processing and optical transportation and so on.

## 2. Simulation results and discussion

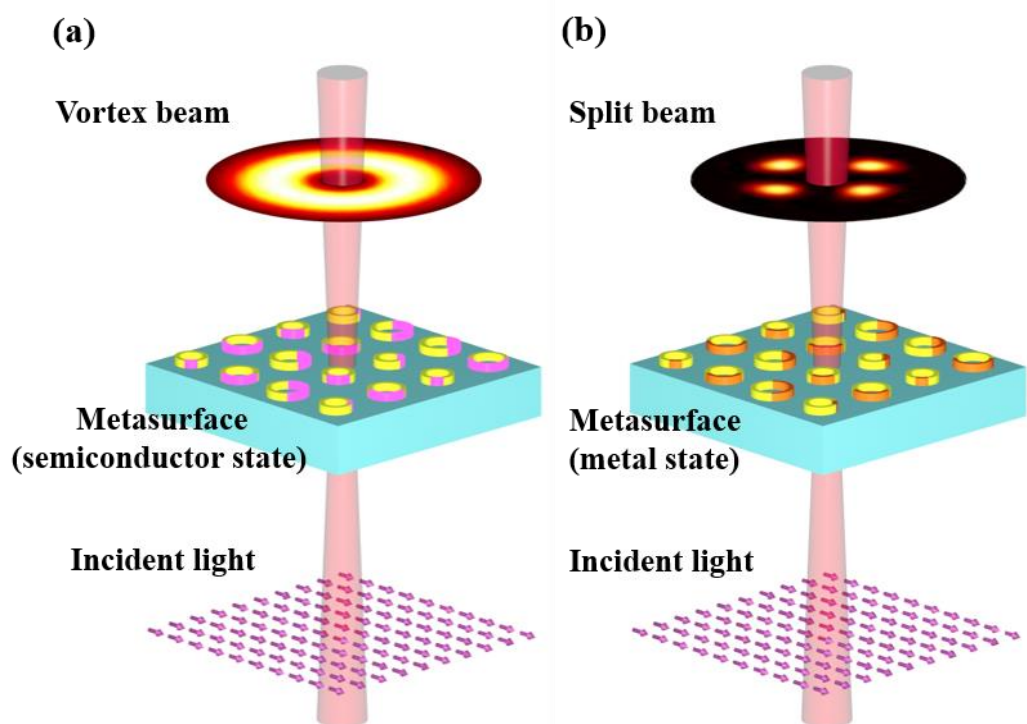

Fig. 1 Schematic of reconfigurable metasurface for dynamic control. Dual functionality can be achieved within a single metasurface for two phase states when illuminated with x-polarization incident light. (a). For the semiconductor state of $VO_2$, such array can produce the vortex beam under the same conditions. (b). For the metal state of $VO_2$, the designed CCRs array can convert the input light into four split beams. The color of purple and orange indicate the semiconductor and metal state of $VO_2$.

The building block of the metasurface (meta-atom) consists of a CCR with different ratio of gold and $VO_2$ on top of a silicon substrate, as shown in Figs. 2(a) and 2(b). Such CCR structures may be fabricated with the state-of-art two step electron beam lithography method. By flexibly varying the inner radius, outer radius, orientation angle and opening angle for the $VO_2$ arc, such geometry can locally modulate the amplitude and phase of the scattered light. In the metal state, the CCR can function similar as a concentric metal ring because the two composite arcs possess similar material properties. In the semiconductor state, the CCR is much more like a split ring wire loop where the loss of $VO_2$ is relatively low, resembling a dielectric gap. The refractive index contrast of $VO_2$ between the two states can result in totally different complex amplitude modulation. When the incident light is polarized along or perpendicular to the symmetry axis of the CCR, there only exist symmetric or antisymmetric mode, respectively. While for arbitrary polarization illumination, $\vec{E}_x^i$ for example, the scattering field $\vec{E}_y^s$ is the superposition of both modes, which can be express as [26]:

$$\vec{E}_y^s = \frac{1}{2}\vec{E}_y^s \sin(2\alpha)(A_s e^{i\Phi_s} + A_{as} e^{i\Phi_{as}}) \quad (1)$$

where $A_s$, $A_{as}$ and $\Phi_s$, $\Phi_{as}$ denote the scattered amplitude and phase from the symmetric and anti-symmetric modes, and $\alpha$ represents the orientation angle, as shown in Fig. 2(b). According to Equation 1, for a fixed geometry of the CCR and $\alpha$ between 0° and 90°, the amplitude of $\vec{E}_y^s$ is solely determined by the orientation angle $\alpha$ and the phase of the scattered field is independent of it. For the cases of $\alpha = \mp 45°$ the absolute value of the amplitude $\vec{E}_y^s$ reaches a maximum polarization conversion. While when $\alpha$ changes from 45° to −45°, the phase of $\vec{E}_y^s$ undergoes another π phase shift, which means one can obtain an additional π phase coverage just by rotating the orientation angle of the CCRs by 90° with uniform scattering amplitude. Hence we achieve simultaneous control of phase and amplitude by such simple design.

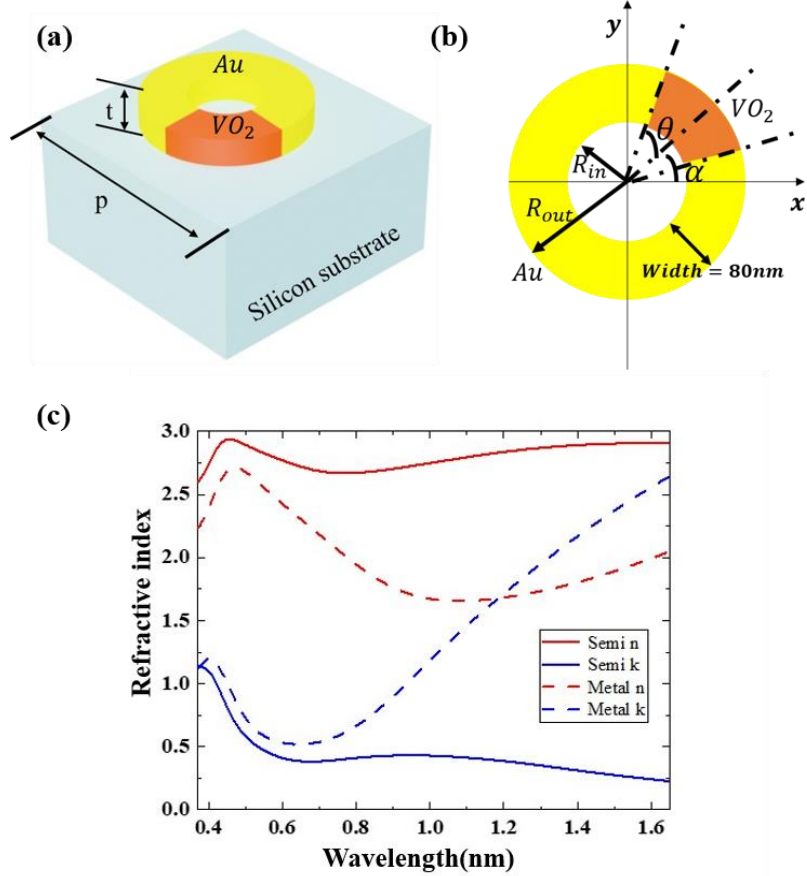

Fig. 2. Design of meta-atom and refractive index of VO$_2$. (a) the meta-atom consists of CCR patterned above silicon substrate. Such CCR composes of different ratio of gold and VO$_2$. t is the thickness of CCR, which is 50nm. p=500nm is the periodicity of the unit cell. (b) The geometry of the meta-atom. $R_{in}$ is the inner radius and the $R_{out}$ is the outer radius of the CCR. $\theta$ is the opening angle while $\alpha$ is the orientation angle. (c) The refractive index of VO$_2$ used in the simulation. Red/blue line corresponding to the semiconductor/metal state and the solid/dashed line is the real/imaginary part of the refractive index. The corresponding data is taken from [38].

To achieve the desired complex amplitude distribution of the transmitted light we carry out a parametric numerical optimization by using a finite difference time domain (FDTD) method. The refractive index of $VO_2$ in the near infrared and visible range is depicted in Fig. 2(c). The refractive index difference between the two states is relatively drastic in the NIR regime, which may benefit the complex amplitude control of the CCRs. The refractive indices of silicon and gold are taken from the data of Palik [50]. Before the implementation of the simulation, we define the duty cycle of the $VO_2$ with $R = \frac{l_{VO_2}}{l_{total}} = \frac{\theta_{VO_2}}{\theta_{total}}$ ranging from 0 to 1 (0 represents no $VO_2$ mixing and 1 corresponds to full $VO_2$ filling), where $l_{VO_2}$, $l_{total}$ are the arc lengths of the $VO_2$ and whole CCR, and $\theta_{VO_2}$ and $\theta_{total}$ are the corresponding center angles, respectively. We conduct a 2D parameter sweep by changing both the outer radius and duty cycle of the $VO_2$ while keeping the width and thickness of CRRs to be 80nm and 50nm respectively. Note that the orientation angle of the CRR is along $45°$ with respect to x coordinate. The incident electromagnetic wave is along the x direction with wavelength of 1500 nm. The periodicity of the unit cell is p=500 nm. The mesh size along x and y direction is both 10nm, and the mesh size in z direction is 20nm. We utilize periodic boundary in both x and y direction. While we set perfect matched layer (PML) boundary condition in z direction. The y component of the transmitted electric field is detected for the analysis. Figs. 3(a) and 3(b) show the simulation results of the transmitted amplitude and phase for the two states of the hybrid CRR, respectively. Based on the calculated data, we select four structures to cover a phase shift of $\pi$ in the semiconductor state and no phase change in the metal state. Each structure possesses almost uniform amplitude modulation. By rotating the symmetry axis with 90°, a full $2\pi$ phase range can be realized. The opening angles for the four CCRs are 130°, 100°, 50°, 60°, the corresponding outer radii are 125 nm, 145 nm, 215 nm, 170 nm, which are also indicated in the hollow rings in Fig. 3 and listed in Table 1.

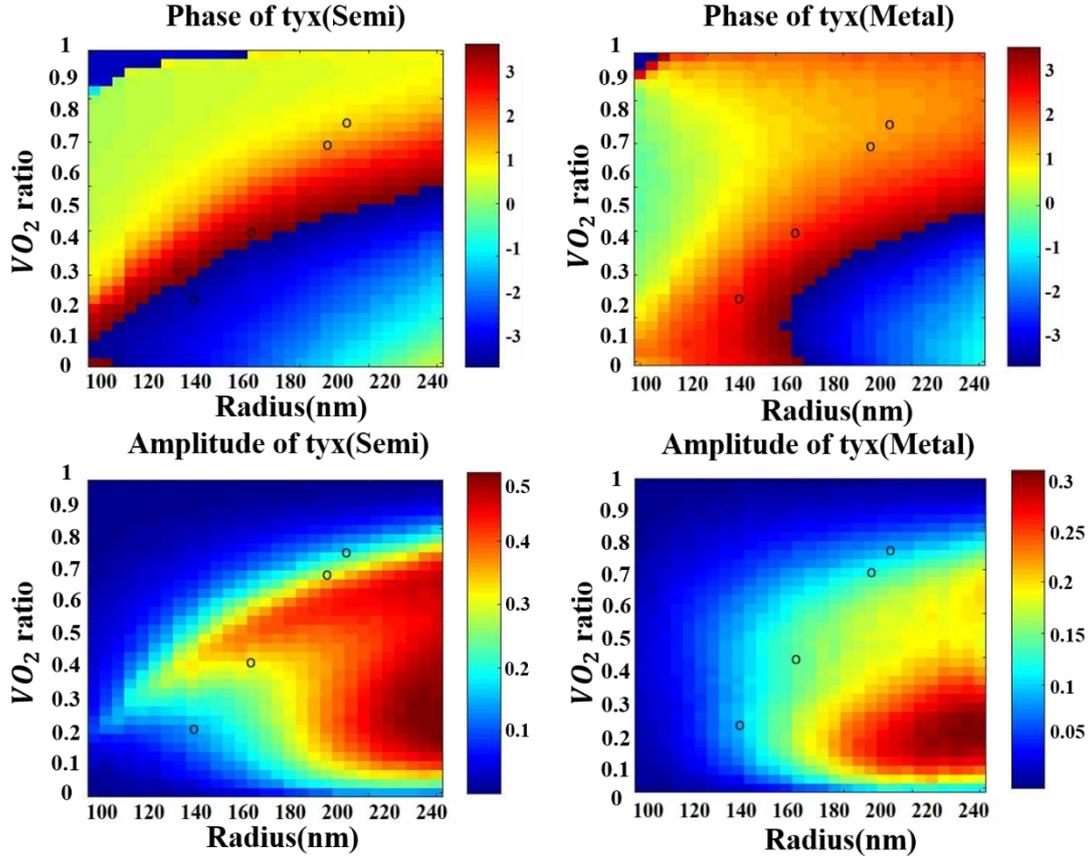

Fig. 3. Simulation results of parameter sweep of CCR. We choose the VO$_2$ ratio ranging from 0 to 1 and outer radius of the rings from 100nm to 240nm, while keeping the width constant at 80nm, to conduct optimization. (a)and(c) corresponding to the phase and amplitude distribution of semi state for the converted y-polarized transmitted light with x-polarized incident wave. Similarly, the results of metal state are shown in (b)and(d). Black hollow circles indicate the selected four structures. Note that tyx means the complex transmitted coefficients of the output transmitted electric field component along y direction with the input incident electric field polarized in x direction.

Table1. Parameters of the selected four CCRs. R$_{out}$ means the outer radius of CCRs with constant width of 80nm.

| Number / Parameter | 1 | 2 | 3 | 4 |
|---|---|---|---|---|
| α | 45° | 45° | 45° | 45° |
| θ | 60° | 50° | 100° | 130° |
| R$_{out}$ | 170nm | 215nm | 145nm | 125nm |

To demonstrate the modulation effect for such hybrid reconfigurable metasurface, for simplicity, we demonstrate anomalous refraction and symmetric diffraction by introducing a phase gradient or phase jump along the interface. The arrangement of the eight structures is illustrated in Fig. 4(a). According to the generalized Snell's law, we have

$$n_t \sin\theta_t - n_i \sin\theta_i = \frac{\lambda}{2\pi}\frac{d\varphi}{dx} \qquad (2)$$

Where $n_i$ and $n_t$ are the refractive index of incident and refractive side correspondingly. $\theta_i$ and $\theta_t$ are the incident and refractive angle, respectively, and $\lambda$ is the incident wavelength in vaccum, $\frac{d\varphi}{dx}$ corresponding to the phase gradient. One can expect that the metasurface in the semiconductor state can bend the light trajectory to an oblique angle of 22.02° under the normal incidence (from substrate to the air). We calculate the electric field and visualize the phase profiles at the x-z plane in both two situations, which are depicted in Fig. 4(b). In the semiconductor state of $VO_2$, the phase profile has a constant phase gradient, resulting in the anomalous refraction. While in the metal state, there exists a staircase-like pattern for the abrupt additional π phase change by rotation of the symmetry axis of the hybrid antennas. The total transmission efficiency at the semiconductor and metal state is 8% and 3%, respectively. We carry out a grating analysis to further explain this phenomenon, as shown in Fig. 4(c). The energy distributions of the different diffraction orders normalized with the total transmission are retrieved. We found that, in the semiconductor state, the energy is mainly directed to the +1 grating order. In contrast, the energy proportion is equivalent in the ±1 order in the metal state.

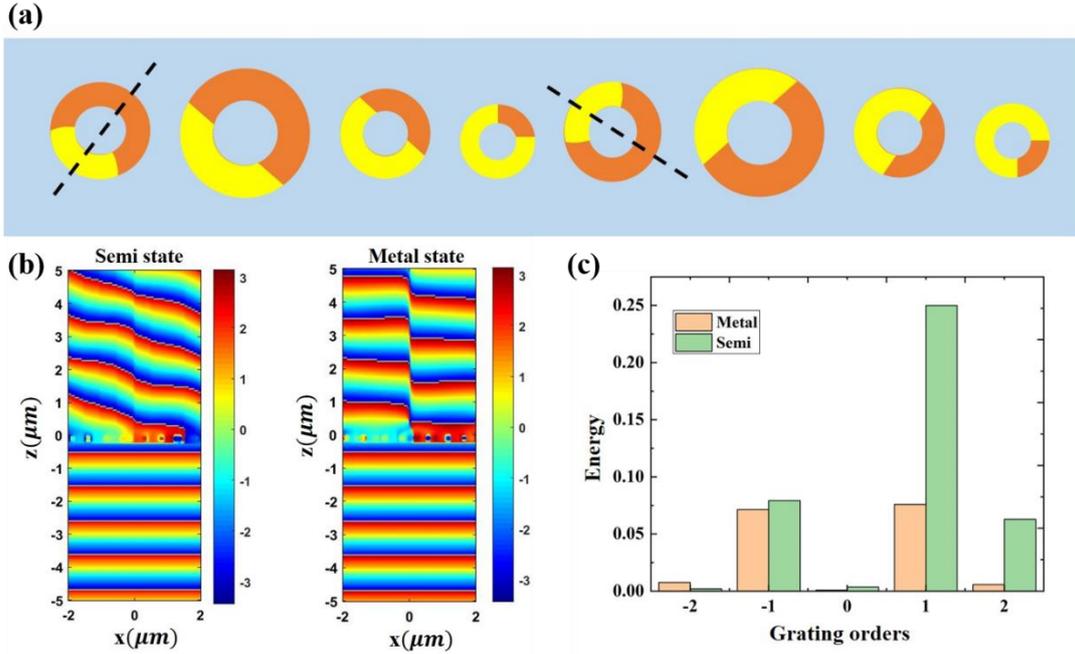

Fig. 4. Anomalous refraction and symmetry diffraction within one metasurface. The incident light with λ=1500nm is linear-polarized along x direction and the recorded output light is along y direction. (a) The top view of the corresponding eight C-shape rings. The black dash lines indicate the orientation angle of the CCRs. The first four CCRs have the same orientation angle of 45° while the angle for the later four CCRs is −45°. (b) the left panel shows the phase distribution corresponding to the semicondutor state where the CCR array can achieve anomalously refraction with oblique angle $\theta_t = \operatorname{asin}\left(\frac{\lambda}{D}\right) = 22.02°$. For the right panel, the phase distribution of the refracted beam shows the staircase-like pattern. (c) The energy distributions of grating orders according to the grating analysis (the ratio of energy distributed to each diffraction orders normalized with the total transmission efficiency), which clearly demonstrates the anomalous refraction and symmetry diffraction phenomenon.

Further, we demonstrate that such reconfigurable metasurface can be utilized to possess a bi-functionality of wavefront engineering. We use the hybrid CCRs array to encode the functionalities for vortex beam generation (mode modulation) and beam splitting (spatial multiplexing) into the two different states. The CCRs array is arranged according to the phase distribution of the vortex beam in semiconductor state. Such a vortex beam possesses the helical phase front characterized by exp($il\varphi$), where $l$ and $\varphi$ are the topological charge and azimuthal angle, respectively. For the selected eight structures in the semiconductor state could cover $2\pi$ phase range while in the same time the phase in the metal state undergoes only 0 and $\pi$ change in different regions. We use an CCRs array with size of 50×50. Figs. 5(a) and 5(b) show the simulation results for the semiconductor state. In Fig. 5(a), the near-field phase distribution is calculated, which has the topological charge of $l$=2 with a phase singularity in the center and indicating a helical phase front. Through the far-field propagation, we observe the donut-shape intensity profile of the vortex beam in Fig. 5(b). As for the metal state as shown in Figs. 5(c) and 5(d), the near-field phase distribution is discontinuous with two areas of 0 and $\pi$, without phase singularity in the center. The far-field intensity distribution of the transmitted beam shows a pattern that is divided into four lobes. In general, such scheme can generate vortex beam with topological charge of $l$=$n$ when in semiconductor states, while the splitting of beams equal to $2n$ in metal states.

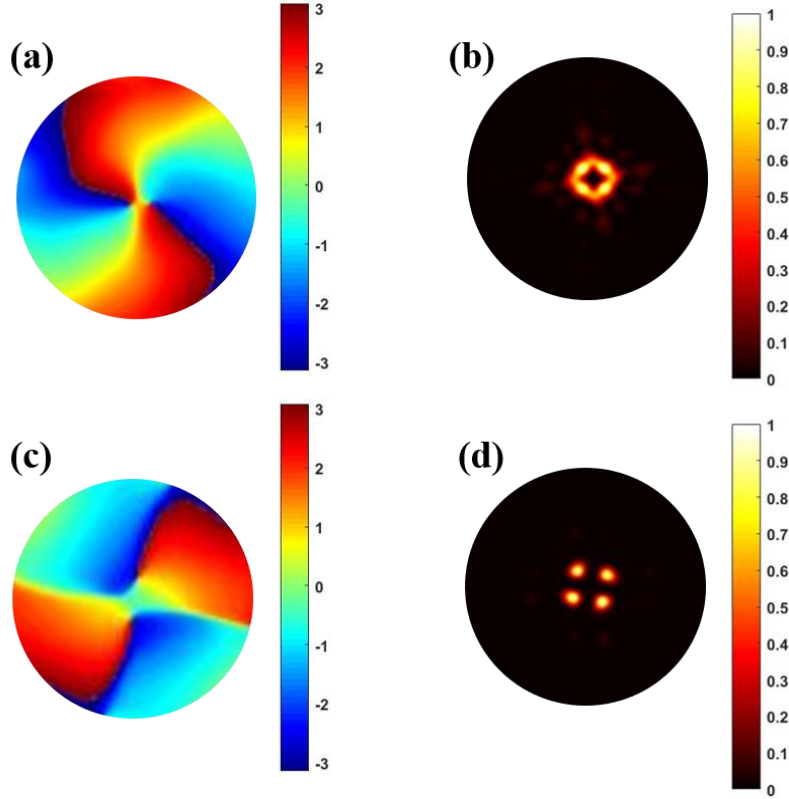

Fig. 5. Simulated results of mode conversion and spatial multiplexing within one metasurface. (a) and (b) are the phase and amplitude distribution of vortex beam generation in semiconductor state. (c) and (d) is the phase and amplitude distribution of four beam splitting in metal state.

Such metasurfaces can also be well exploited in fast sensing and optical manipulation applications. In the metal state of our metasurface, the incident light beam can become quadruplex which can be used for parallel fabrication and detecting/sensing. While the output vortex beam can be applied as optical tweezer for manipulating biomolecules, nanoparticles and so on, under the semiconductor state. Due to the less contrast of refractive indices between the intermediate states may result in less modulation depth, that is, the phase modulation of the orthogonal component tyx is shrink to less than 0 to π range compared to the semiconductor state, and get closer to a flat modulation as the metal state. Hence, in the intermediate states varied from completely semiconductor state to completely metal state, the output light can still gradually change from vortex beam to four petals, with the topological charge evolution from integer to fractional value and finally mode splitting. As the switching speed between two states of $VO_2$ can be as fast as picoseconds, the dynamic change of the functionality could be even used for relatively fast processes. Furthermore, the low-temperature threshold required for the phase transition between two states of $VO_2$ makes such metasurface be promising for low-loss integrated photonic circuits, especially for the case when the operating temperature typically cannot exceed one hundred degree Celsius in thermal stimulus. We can further improve the efficiency by designing Metal-Insulator-Metal (MIM) scheme for the reflective type, which may enhance the efficiency while possessing the same complex amplitude modulation properties at different phase change states. In comparison with commercial beam splitters/spiral phase plates reported, such hybrid reconfigurable metasurface is optical thin, easy to fabricate, and compatible with flat optics, more importantly, it can integrate two functions into one element.

### 3. Conclusions

In summary, we have demonstrated the complex amplitude modulation of a hybrid metasurface composed of nanostructure made of a dynamically reconfigurable phase change material. By building composite concentric ring structures with different ratio of gold and $VO_2$, we achieve uniform amplitude and $2\pi$ phase shift coverage in both metal and semiconductor states. With the realizing of anomalous refraction resulting from a phase gradient in the semiconductor state and symmetric diffraction resulting from a step-like phase profile in the metal state, we demonstrate that such metasurface can implement dual functionalities. That is, dynamic control can be achieved by changing the environmental temperature or by external femtosecond laser pulse stimulus. Furthermore, we have demonstrated switchable vortex beam generation and beam splitting phenomenon. Compared to previously designed metadevices working in the far/mid-infrared regime, such metasurfaces can pave the way for applications in the NIR band and even have the potential to extend to the optical regime. Importantly, such metadevices may serve as novel platforms for parallel laser fabrications, signal processing, memory storage, holography, anti-counterfeiting and so on.

**Funding**

The authors acknowledge the support from the National Natural Science Foundation of China (No. 61775019) program, the Beijing Municipal Natural Science Foundation (No. 4172057), the Beijing Nova Program (No. Z171100001117047), the Fok Ying-Tong Education Foundation of China (No.161009). We also acknowledge the NSFC-DFG joint program (DFG No. ZE953/11-1, NSFC No. 61861136010).